\documentclass[preprint,showkeys,preprintnumbers,amsmath,amssymb]{revtex4}

\usepackage{verbatim}

\usepackage{graphicx}
\usepackage{dcolumn}
\usepackage{bm}

\newcommand{\bra}[1]{\langle #1|}
\newcommand{\ket}[1]{|#1\rangle}

\newcommand{\expect}[1]{\langle #1 \rangle}

\newcommand{\nuoli}[1]{\rightarrow}
\newcommand{\abs}[1]{\left| #1 \right|}
\newcommand{\omaexp}[1]{ \exp \left( #1 \right)}
\newcommand{\omalog}[1]{ \ln \left( #1 \right)}

\begin{document}

\title{Polymer dynamics in time-dependent periodic potentials}

\author{Janne Kauttonen}
 \affiliation{Department of Physics, University of Jyv\"askyl\"a, P.O. Box 35, FI-40014 Jyv\"askyl\"a, Finland}
 \email{janne.kauttonen@phys.jyu.fi}
\author{Juha Merikoski}
 \affiliation{Department of Physics, University of Jyv\"askyl\"a, P.O. Box 35, FI-40014 Jyv\"askyl\"a, Finland}  
\author{Otto Pulkkinen}
 \affiliation{Institut f\"ur Theoretische Physik, Universit\"at zu K\"oln, Z\"ulpicherstr. 77, 50937 K\"oln, Germany}
  
\date{\today}

\begin{abstract}
Dynamics of a discrete polymer in time-dependent external potentials is studied with the master equation approach.  We consider 
both stochastic and deterministic switching mechanisms for the potential states and give the essential equations for computing the 
stationary state properties of molecules with internal structure in time-dependent periodic potentials on a lattice.  As an 
example, we consider standard and modified Rubinstein-Duke polymers and calculate their mean drift and effective diffusion 
coefficient in the two-state non-symmetric flashing potential and symmetric traveling potential.  Rich non-linear behavior of 
these observables is found.  By varying the polymer length, we find current inversions caused by the rebound effect that is only 
present for molecules with internal structure.  These results depend strongly on the polymer type.  We also notice increased 
transport coherence for longer polymers.

\end{abstract}

\maketitle

\section{Introduction}

There has been considerable progress in the research of Brownian motors during the last decade (see {\it 
e.g.}~\cite{Reimann,Linke,Astumian}).  Starting with the simple pointlike Brownian particles with time-dependent driving forces, 
research has expanded towards the more complex objects such as interacting Brownian particles ({\it 
e.g.}~\cite{Wang,Klumpp,Chen,Gehlen,Fendrik,Retkute}) and polymers \cite{Craig1,Craig2,Streek}.  In this paper, we study polymer 
motion with discrete lattice models, which allows us to consider different kinds of microscopic polymer dynamics in detail.  Aside 
from being a purely theoretical branch of study, analysis of simplified discrete non-equilibrium particle models has became an 
important tool for studying biologically inspired Brownian motor systems ({\it e.g.}~\cite{Fisher}).

Discrete models have been applied widely to single particle ratchet problems ({\it 
e.g.}~\cite{Jarzynski,Schimansky,Pascual,Zhou}).  We expand this picture by considering a generalized Rubinstein-Duke model (RD 
model) \cite{Rubinstein,Duke} for polymer motion in discrete time-dependent potentials.  An interesting question is, what kind of 
dynamics lies beyond simple pointlike particles and how one can calculate its properties such as the effective diffusion 
coefficient and the drift.  Although there are plenty of studies concerning the behavior of the RD polymer in zero (or uniform) 
field ({\it e.g.}~\cite{Widom,Carlon}), only recently a ratchet mechanism (tilting ratchet) has been considered in this context 
\cite{Puola2}.

Especially because of the high complexity of Brownian motors with internal structure, most studies of these systems have applied 
the Monte Carlo method.  However, since the ratchet systems are quite sensitive to the values of parameters, and drifts generated 
by pure ratchet mechanism are usually very small, Monte Carlo simulations tend to be very time-consuming and inaccurate.  In this 
paper we study these systems with the master equation approach.  The results obtained this way are accurate enough to reveal the 
details of the dynamics.

The purpose of this paper is to give a hands-on example of how one applies the master equation method to systems involving 
time-dependent periodic potentials and complex molecules by using a modified RD model polymer as a prototype of such molecules.  
We perform calculations for short linear polymers in a non-equilibrium environment generated by flashing and traveling ratchets.  
To test the significance of the polymer type, reptating or not, we compare the motion of the RD polymer with the dynamics of a 
modified version of the RD polymer with less constrained microscopic movement.

The paper is organized as follows: In Section II we expand the modified RD polymer model to periodic time-dependent potentials and 
give equations for the calculation of the drift and diffusion coefficients, in Section III we present results of the calculations 
for short polymers and finally in Section IV we give our conclusions and discuss the implications to applications.

\section{The Model and methods}

The RD model was originally developed to model the random motion of a flexible polymer in a confined medium with static obstacles 
({\it e.g.}~pores in gel) that the polymer must bypass, therefore causing the polymer {\it reptation}.  By assuming that the 
network of obstacles can be modeled (on average) by a lattice-like structure, that the correlation length between the polymer 
segments is smaller than the distance between the obstacles, and that only the polymer heads are able to move into previously 
unoccupied cells (lattice sites), the problem can be discretized to a simple particle hopping model \cite{Rubinstein}.  Soon after 
the original model was expanded \cite{Duke} to be suitable for external potentials ({\it e.g.}~static field), pure theoretical 
research of the model started to flourish such as in Refs.~\cite{Widom,Kooiman}.

Technically the RD model is a spin-1 chain with special kind of nearest neighbor interactions between the particles ({\it 
reptons}).  By assuming that the reptons experience random "pushes" by the environment modelled with a continuous time Markov 
process with exponentially distributed waiting times, we can construct the stochastic generator of the system.

In order to study the effect of the intrinsic transition rules of the polymer in time-dependent periodic potentials on long-time 
dynamics, we will compare the results of the RD model to the results of a non-reptating
polymer which allow the breaking of the reptation tube.  In this paper we call this extended model the \emph{free-motion} model 
(FM model).  In Fig.~1 there is an illustration of an example configuration of a six repton polymer with arrows indicating all 
allowed moves for both RD and FM models (see also Ref.~\cite{constrait3}).  All repton transitions are between nearest neighbor 
lattice sites only.  Similar extensions have been studied previously in a different context in 
Refs.~\cite{constrait3,constrait1,constrait2}.

\begin{figure}
\includegraphics[width=6.0cm]{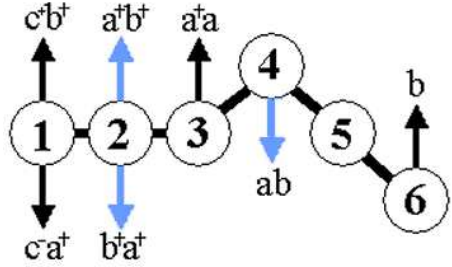}
\caption{(Color online) Illustration of the allowed transitions in RD and FM models for a six repton polymer in one of its 
configurations.  The moves described by the blue arrows are only allowed for the FM polymers and those by the black arrows for 
both polymer types.  The letters a, b, c represent the operators corresponding to the moves and are defined later in the text.}
\end{figure}  

As an environment for the polymers, we assume a discrete periodic potential $V(x)$ such that
\begin{equation*}
V(x+L)=V(x) .
\end{equation*}
To make contact with Kramers rate theory (see {\it e.g.}~review \cite{Kramers}) and the previous work related to discrete ratchets 
\cite{Schimansky}, we define the transition rate from state $i$ to $j$ by
\begin{gather*}
\gamma_{i \rightarrow j} = \alpha \omaexp{\beta \left( V(i) - V(j) \right)} ,
\end{gather*}
and choose $\alpha = \beta = 1$ to define the time and energy scales along with the lattice constant $1$ to define the spatial 
length scale.  We shall next define the time-evolution operators for the RD and FM models (readers not interested in the formal 
development may skip the rest of this section).

The mathematical model for the polymer, which contains the RD model as a special case but also allows breaking of the reptation 
tube if wanted, is constructed as follows (see {\it e.g.}~\cite{Sartoni}).  Within the most compact, the inner coordinate 
representation, every bond between reptons can be in three states; up (state A), down (state B) or flat (state 0).  In Fig.~1, 
reptons 1, 2 and 3 are in state 0, repton 4 in state A, and reptons 5 and 6 in state B.  An $N$-repton polymer has $N-1$ bonds.  
The state corresponding to polymer configuration $y$ is thus given by a $3^{N-1}$-dimensional state vector $\ket{ \Psi_{y} }$.

The non-zero elements of the local creation and annihilation operators defining the dynamics of the bonds are
\begin{align*}
\left[ n_A \right]_{1,1} &= \left[ n_0 \right]_{2,2} = \left[ n_B \right]_{3,3} = 1 \\ 
\left[ a \right]_{2,1} &= \left[ a^{\dagger} \right]_{1,2} = \left[ b \right]_{2,3} = \left[ b^{\dagger} \right]_{3,2}  = 1 .
\end{align*}
The operators $a$ and $b$ produce changes in the local bond configuration as indicated in Fig.~1.  To extend the model to include 
a periodic potential $V$, we must add an additional state.  One repton is chosen as a marker repton that keeps track of the 
polymer position within the potential.  The transition rates of a single repton now depend on the position of the marker repton 
and all other bonds separating it from the marker.  Either of the head reptons is the most convenient choice for the marker 
repton, hence we choose here the repton labeled $1$ (see Fig.~1).  The dimension of the marker state is $L$, so the dimension of 
the total system of equations becomes $L \times 3^{N-1}$.  

By denoting
\begin{align*}
L(i)&= \omaexp{-V(i+1)+V(i)} \\
R(i)&=\omaexp{-V(i-1)+V(i)} ,
\end{align*}
indicating transitions to left and right (corresponding down and up in Fig.~1), the non-zero matrix elements for the marker state 
and transition operators are
\begin{gather*}
\left[ c_l^{-} \right]_{l-1,l} = 1 \text{ for $l \neq 1$}, \quad \left[ c_{l}^{+} \right]_{l+1,l} = 1 \text{ for $l \neq L$} \\
\left[ c_1^{-} \right]_{L,1} = \left[ c_{L}^{+} \right]_{1,L} = \left[ n_l \right]_{l,l} = 1 ,
\end{gather*}
where $1 \leq l \leq L$.  The state of the polymer now has the form
\begin{equation*}
\ket{\text{marker repton}} \otimes \ket{\text{polymer configuration}}=
\ket{\Psi_{l}} \otimes  \ket{\Psi_{y}} ,
\end{equation*}
where $\ket{ \Psi_{l} }$ is the marker repton state vector with dimension $L$.  The stochastic generator of the polymer model in 
the $L$-periodic potential thus becomes
\begin{equation}
H = \sum_{l=1}^{L} \left[ A_{l} + \sum_{y} \left( B_{y,l}+\sum_{i=1}^{N-2} M_{i,y,l} \right) \right] ,
\label{eq:Hinner}
\end{equation}
where the operator $A$ applies to bond $1$ and the marker repton, $M$ applies to bulk reptons and $B$ applies to bond $N-1$.  The 
explicit forms of these operators are given in Appendix A.  

\subsection{Time-dependent potentials}
The time-dependence of the environment can break the detailed balance and may result in a directed drift.  We assume that the 
switching between the distinct environments is independent of the polymer state in the potential {\it i.e.}~there is no feedback 
from the polymer.  The switching mechanism between the potentials can be either stochastic or deterministic.  The stochastic 
Markovian switching allows us to evaluate the stationary state directly by solving an eigenvalue problem.  This is the most widely 
used way of studying Brownian motors and similar systems.  With deterministic switching, we must numerically integrate to get the 
periodically stationary state.

Due to the time- and position-dependent transition rates, extra care must be taken to numerically study the process accurately.  
For example, one should not use the standard discrete-time Monte Carlo simulation method that has been widely used in various RD 
model studies.  It does not produce correct results for our models.  Instead one should handle the master equation directly by 
means of numerical integration or use the continuous-time Monte Carlo method.  In general, transport of particles in 
time-dependent potentials is a hard problem to solve exactly.  Even for a single particle in a periodic potential the general 
solution is not known.  The solution for stationary potentials, however, is available (pioneered by Derrida \cite{Derrida}).

First assume Markovian switching between the potentials.  We must include an additional state that keeps track of the current 
potential
\begin{multline*}
\ket{\text{potential state}} \otimes \ket{\text{marker repton}} \otimes \ket{\text{polymer configuration}}
=\ket{\Psi_{s}} \otimes \ket{\Psi_{l}} \otimes  \ket{\Psi_{y}} ,
\end{multline*}
where $\ket{\Psi_{s}}$ is the state vector of the potential with dimension $S$ {\it i.e.}~the number of different potentials.  
Since there is no feedback mechanism, adding this new state is straightforward.  The non-zero state and transition operator 
elements for the potential state are
\begin{align*}
\left[ \hat{h}_{i} \right]_{i+1,i} &= 1 \text{ for $1 \leq i \leq S$ } & \left[ \hat{h}_{S} \right]_{1,S} = \left[ n_s 
\right]_{s,s} = 1, 
\end{align*}
where $1 \leq s \leq S$.  By defining the operator $\hat{h}$ like this, we consider only cyclic transitions between the potentials 
({\it i.e.}~$1 \rightarrow 2 \rightarrow \dots \rightarrow S \rightarrow 1 \rightarrow \dots$) to preserve the analogy with the 
deterministically switching potentials.  The stochastic generator becomes
\begin{equation*}
H = \sum_{s=1}^{S} \left[ \tilde{H}_{\rm s} + T_{s}^{-1} \left( n_s - \hat{h}_{s} \right) \right] ,
\end{equation*}
where $\tilde{H}_{\rm s}$'s are formed by extending all elements of the operator in Eq.~(\ref{eq:Hinner}) with their corresponding 
potential state $s$ ({\it e.g} \; $a_{i,j} \rightarrow n_s a_{i,j}$) and $T_1,T_2,\dots,T_S$ are the mean life-times of the 
potentials.

With deterministic switching, the stochastic generator is given by
\begin{equation*}
H(t)=
\begin{cases}
H_{\rm 1} \: , \quad t \in \left[ 0,T_1 \right)
\\ H_{\rm 2} \: , \quad t \in \left[ T_1,T_1+T_2 \right)
\\ \vdots
\\ H_{\rm S} \: , \quad t \in \left[ \sum_{i=1}^{S-1} T_i,T \right) ,
\end{cases}
\end{equation*}
where $H_{\rm s}$ is the operator of the type (\ref{eq:Hinner}) in the potential $s$ and $T=\sum_{i=1}^{S} T_i$ is the time-period 
and $H(t+T)=H(t)$.  In this case there exists a $T$-periodic stationary solution.  Once $H$ is given, the time-evolution of the 
system is governed by the master equation $d q(t)/dt = H(t) q(t)$, where $q(t)$ is the probability vector.  Since $H_{\rm s}$'s 
are generally non-symmetric, $q(t)$ usually has an oscillating behavior.

\subsection{Drift and diffusion}

We are interested in the drift and diffusion of the center of mass of the polymer.  The velocity and the diffusion coefficient can 
be defined as
\begin{align*}
v &= \lim_{t \rightarrow \infty} \frac{d}{dt} \expect{x_{\rm CM}(t)} \\ 
D_{\rm eff} &= \frac{1}{2} \lim_{t \rightarrow \infty} \frac{d}{dt} \left( \expect{x_{\rm CM}(t)^2} -  \expect{x_{\rm CM}(t)}^2 
\right) ,
\end{align*}
where $x_{\rm CM}$ is the center of mass of the polymer.  Here $v$ and $D_{\rm eff}$ could also be defined for single reptons 
instead of the center of mass and this local approach naturally leads to the same longtime values.  

From the previous we define the Peclet number
\begin{gather*}
\text{Pe} =  \frac{ \abs{ v \: \ell }}{D_{\rm eff}} ,
\end{gather*}
where we choose the length scale $\ell = 1$.  Since our polymer is simply composed of several neighbor-hopping random walkers, we 
can generalize the formalism of Ref.~\cite{Pascual} (which generalizes the ideas of Ref.~\cite{Derrida}).  First define
\begin{align}
q_{y}(t) &= \sum_{n=-\infty}^{\infty} p_{n,l,y'}(t) \label{eq:differentioi1} \\
s_{y}(t) &= \sum_{n=-\infty}^{\infty}(l+nL) p_{n,l,y'}(t) - \expect{x_{\rm CM}(t)}q_{y}(t) , 
\label{eq:differentioi2} 
\end{align}
where $p_{n,l,y}$ is the probability to find the marker-repton in the position $l + n L$ with the polymer inner configuration $y'$ 
and the re-defined state $y$ includes both the marker-repton position ($l$) and the inner configuration ($y'$) within the 
L-periodic potential.  Assume that the stochastic generator $H$ of the total system is defined by the rates $\Gamma_{i,j}$ from 
state $i$ to $j$.  It can be shown by using the definitions above, by taking the time-derivatives and using the master equation 
(see {\it e.g.} \cite{Kolomeisky} for a similar calculation) that
\begin{gather*}
v(t)=h \sum_{y} \left( \overline{R}_{\rm out}^y - \overline{L}_{\rm out}^y \right) q_{y}(t) ,
\end{gather*}
where
\begin{gather*}
\overline{R}_{\rm out}^y = \sum_{i}^{\rightarrow} \Gamma_{y,i} \quad \quad 
\overline{L}_{\rm out}^y = \sum_{i}^{\leftarrow} \Gamma_{y,i} \; 
\end{gather*}
with arrows indicating the direction (right or left) of those repton transitions that lead from the state $y$ to states $i$, 
neglecting all the rest.  Since this expression is for the center of mass, $h=1/N$ is chosen as the new lattice constant.  

Similarly we get
\begin{gather*}
D_{\rm eff}(t) = \frac{h^2}{2} \sum_{y} \left( \overline{R}_{\rm out}^y + \overline{L}_{\rm out}^y \right) q_{y}(t) \\
+ h \sum_{y} \left( \overline{R}_{\rm out}^y - \overline{L}_{\rm out}^y \right) s_{y}(t) . 
\end{gather*}
The evolution equations for $q_{y}(t)$ and $s_{y}(t)$ can be found by differentiating (\ref{eq:differentioi1}) and 
(\ref{eq:differentioi2}) in time and using the master equation once more.  We arrive at
\begin{gather*}
\frac{d q_y(t)}{dt}= - \left( \overline{R}_{\rm out}^y + \overline{L}_{\rm out}^y \right) q_y (t) + \overline{R}_{\rm in}^{y} 
(q(t)) + \overline{L}_{\rm in}^{y} (q(t)) \\
\frac{d s_y(t)}{dt}= - \left( \overline{R}_{\rm out}^y + \overline{L}_{\rm out}^y \right) s_y (t) + \overline{R}_{\rm in}^{y} 
(s(t)) + \overline{L}_{\rm in}^{y} (s(t)) \\ - h \left[ \overline{L}_{\rm in}^{y}(q(t)) - \overline{R}_{\rm in}^{y}(q(t)) \right] 
- v(t) q_y(t) ,
\end{gather*}
where
\begin{align*}
\overline{R}_{\rm in}^{y}(q(t)) &= \sum_{i}^{\rightarrow} \Gamma_{i,y} q_i(t) \: \quad \overline{R}_{\rm in}^{y}(s(t)) = 
\sum_{i}^{\rightarrow} \Gamma_{i,y} s_i(t)  \\
\overline{L}_{\rm in}^{y}(q(t)) &= \sum_{i}^{\leftarrow} \Gamma_{i,y} q_i(t) \: \quad
\overline{L}_{\rm in}^{y}(s(t)) = \sum_{i}^{\leftarrow} \Gamma_{i,y} s_i(t) \: .
\end{align*}

Note that all transitions are assumed to be between nearest neighbor lattice sites only.  Otherwise transitions of certain length 
should be collected in their own sets according to their hopping distances ($ \leq L$), which would appear as coefficients of 
additional sum-terms in the equations.  In the matrix form
\begin{align}
\frac{d q(t)}{dt} &= H(t) q(t) \label{eq:matrixform1} \\
\frac{d s(t)}{dt} &= H(t) s(t) - h H_{\rm sign}(t) q(t) - v(t) q(t) \label{eq:matrixform2},
\end{align}
where $H_{\rm sign}$ has the structure
\begin{align*}
\left[ H_{\rm sign} \right]_{i,j} &= \left[ H \right]_{i,j} \text{for all right transitions} \\
\left[ H_{\rm sign} \right]_{i,j} &= -\left[ H \right]_{i,j} \text{for all left transitions} \\
\left[ H_{\rm sign} \right]_{i,j} &= 0 \text{ for all other transitions } i,j \; .
\end{align*}
This operator is easily built while building the stochastic generator itself.  Since $v(t)$ in Eq.~(\ref{eq:matrixform2}) is 
governed by Eq.~(\ref{eq:matrixform1}), these systems must be solved simultaneously.  See also Refs.~\cite{Maritan,Elston} where 
similar approach has been applied to find the drift and the effective diffusion coefficient for complex molecules.

\subsubsection{Time-independent stationary states}

When $H$ is time-independent, we can take the limit ${t \rightarrow \infty}$ and define the steady-state parameters as
\begin{align*}
Q_{y} &= \lim_{t \rightarrow \infty} q_{y}(t) & S_{y} &= \lim_{t \rightarrow \infty} s_{y}(t) . 
\end{align*}
By using these we get well-defined stationary values
\begin{gather}
v = h \sum_{y} \left( \overline{R}_{\rm out}^y - \overline{L}_{\rm out}^y \right) Q_{y} \notag \\
D_{\rm eff} = \frac{h^2}{2} \sum_{y} \left( \overline{R}_{\rm out}^y + \overline{L}_{\rm out}^y \right) Q_{y} \notag \\ 
+ h \sum_{y} \left( \overline{R}_{\rm out}^y - \overline{L}_{\rm out}^y \right) S_{y} 
\label{eq:statdiff}
\end{gather}
for the velocity and the effective diffusion coefficient.  Now $Q_{y}$'s and $S_{y}$'s are found by solving the equations
\begin{align}
H Q &= 0 &  H S &= h H_{\rm sign} Q + v Q .
\label{eq:nonhomogeneous}
\end{align}
So far equations like these have been solved exactly only for a single particle on a periodic lattice.  The first solution was 
given in Ref.~\cite{Derrida} for the nearest neighbor hopping particle with arbitrary transition rates.  This has been later 
extended {\it e.g.}~for parallel one-dimensional lattices in Ref.~\cite{Kolomeisky}.  However, for more complex systems (like RD 
polymers), solutions cannot be found by exact methods and numerics must be applied.  The structure of $H$ also raises some issues.  
Since the determinant of $H$ is always zero, mathematically there is no unique solution for the non-homogeneous linear set of 
equations in (\ref{eq:nonhomogeneous}).  This can be easily seen by using the fact that an ergodic stochastic system always has a 
non-trivial stationary state, therefore by the \emph{rank-nullity theorem} we have $\text{Rank}(H) = \text{Dim}(H)-1$, meaning 
that we have one free parameter and all solutions are separated by a constant ({\it i.e.}~$S_{y}$ is a solution $\Leftrightarrow$ 
($S_{y}$ + constant) is a solution).  However, since we also have conditions
\begin{align*}
\sum_{y} Q_y &= 1 & \sum_{y} S_y &= 0 ,
\end{align*}
which can be derived from the definitions of $s_y(t)$ and $q_y(t)$, there indeed exist unique solutions for $S$ and $Q$ (which is 
of course also required on physical grounds).

Eq.~(\ref{eq:statdiff}) is a generalization of the result derived in Ref.~\cite{Widom}.  This can be seen by considering the case 
$L=1$ without external potentials ({\it i.e.}~$y$'s are simply inner configurations, $v=0$ and $Q := Q_y=3^{-N+1}$), so that with 
$a_y := -2 S_{y}/hQ$ we have (for lattice constant $1$)
\begin{gather*}
D_{\rm eff} = \frac{h^2}{2} Q \sum_{y} \left[ \left( \overline{R}_{\rm out}^y + \overline{L}_{\rm out}^y \right) + \frac{2}{h Q} 
\left( \overline{R}_{\rm out}^y -  \overline{L}_{\rm out}^y \right) S_{y} \right] \\
= \frac{1}{2N^2} \frac{1}{3^{N-1}} \sum_{y} \left[ \left( r_y + l_y \right) - \left( r_y  - l_y \right) a_{y} \right] .
\end{gather*}
Here we used the fact that, in this case, every state $y$ has a weight $1/3^{N-1}$ and $r_y/l_y$'s can be interpreted as the 
"number of arrows" for right/left transitions out from the state $y$.  In Ref.~\cite{Widom}, symmetry properties (reflections) of 
polymer configurations were used to find a unique solution for $a_y$'s, but this is not possible when external potentials are 
present and the problem is non-symmetric.  However, numerical linear algebra tools can be used to find the solution.     

\subsubsection{Time-dependent stationary states}

When $H$ is time-dependent, we must integrate equations (\ref{eq:matrixform1}) and (\ref{eq:matrixform2}) in time until the system 
arrives at the periodically stationary state (with period $T$).  The mean velocity and the diffusion coefficient are determined by
\begin{align*}
v &= \lim_{t \to \infty} \frac{1}{T} \int_{t-T}^t v(s) \, ds  \\
D_{\rm eff} &= \lim_{t \to \infty} \frac{1}{T} \int_{t-T}^t D_{\rm eff}(s) \, ds .
\end{align*}
In practice, these are calculated by integrating in time long enough so that results have converged.

\subsection{Fast and slow switching regimes}

When the switching times of the potential are close to the characteristic timescales of the system ({\it i.e.}~relaxation times), 
the behavior depends heavily on the switching type and lifetimes of the states.  However, when the potential changes very rarely 
or extremely fast, the system becomes independent of the switching type and even of the relative life-times of the states.

First assume that the total mean switching period $T \rightarrow 0$ such that $T_i > 0$ for all mean lifetimes of the potentials 
($1 \leq i \leq S$).  In this case particles experience an effective average potential ("mean-field" \footnote{Not to be mixed up 
with the mean-field in the sense of an approximation tool for the stochastic generator $H$ itself}) and the transition rates 
become
\begin{equation}
\Gamma_{i,j}^{\rm MF} = \sum_{k} x_k \Gamma_{i,j}^{k} ,
\label{eq:meanfield} 
\end{equation}
where $\Gamma_{i,j}^{k}$ are the transition rates of the stochastic generator of type (\ref{eq:Hinner}) in the potential $k$ and 
$x_k = T_{k}/T$'s are weight factors determined by the mean life-times of the potentials.  This leads to a mean-field stochastic 
generator with dimension $L \times 3^{N-1}$.  This approach was used in Ref.~\cite{Stukalin} to solve exactly the single particle 
dynamics in two arbitrary alternating periodic potentials.  Although this mean-field limit is mathematically well defined, from 
the physical point of view it's artificial since real-world systems have inertia, and changing the potential state takes some 
finite time ({\it e.g.}~charge re-distribution to build up an electric field).  So the velocity always goes to zero in the fast 
switching limit.

Now assume that $T_i \gg \tau_i$ for all $1 \leq i \leq S$ where $\tau_i$ is the longest relaxation time of the system in the 
potential $i$.  This means that the system always converges close to the stationary state in the current potential before the 
potential is switched to the next one.  By the model assumptions, drift is always zero at the stationary state in all potentials.  
Let $d_{j|i}$ denote the mean travel distance of the molecule center of mass within the potential $j$ using the stationary state 
of the potential $i$ as an initial state and then letting the system fully relax \footnote{Note that when $V$ is relatively small, 
the integral $\lim_{t \rightarrow \infty} \int_{0}^{t} v(s) ds \,$ dies out fast and in practice we can truncate the integral at 
some moderate value of $t$ (i.e.~few times of the relaxation time).}.  Summing over all $d_{j | i}$'s gives the total expected 
distance within one time-period $T$, and by assuming cyclic switching of the states, we define
\begin{equation}
d=\sum_{i=1}^{S} d_{i | i+1} . 
\label{eq:slowlimit} 
\end{equation}
The sign of $d$ determines the drift direction in the large $T$ limit and the asymptotic drift thus becomes $v = d/T$.

That internal molecular states may have strong influence on the dynamics can be already seen in the slow switching regime.  
Letting the molecule first find its equilibrium in some non-flat potential and then turning the potential off may indeed result in 
directed motion of the molecule after the switching, due to internal relaxation, whereas a single particle would be immobile in 
the mean.  These rebounds might be dominating and define the sign of $d$.

\section{Results}

\subsection{Choice of the potentials}

We have numerically analyzed RD and FM models with the polymer length of $N=1...11$ reptons and with two potentials ($S=2$) and 
stochastic switching.  All calculations were done with MATLAB.  We used the standard Runge-Kutta 4 method to integrate 
(\ref{eq:matrixform1}) in time and a trapezoid method to calculate the resulting integral in (\ref{eq:slowlimit}).  The Arnoldi 
and BiConjugate gradient stabilized methods were used to solve homogeneous and non-homogeneous systems in 
(\ref{eq:nonhomogeneous}).

We consider two basic potential types: flashing and traveling ratchets.  The first type is the most general non-symmetric 
potential that has been extensively used in studies of Brownian motors and the latter one is a generic example of asymmetrically 
placed symmetric potentials and has been recently used with single particle models \cite{Grier,Pascual}.  We consider the simplest 
case $L=3$, which is the smallest possible length that can form both of these potentials with the ratchet effect.  See Fig.~2 for 
sketches of these potentials.  A positive drift sign indicates motion in the increasing lattice-site index direction.  Because of 
symmetries, the next choice would be $L=5$, but this choice would also need longer polymer lengths ($N \gg 11$) than we can 
efficiently handle.  We require that the polymer must be able to cover several potential periods when fully extended.  We also set 
$V_{\rm max}=1/2$, which we found to give interesting results while also being computationally feasible \footnote{Since the 
asymmetry of the stochastic generator and the relaxation time of the polymer grow exponentially as the potential becomes larger, 
the problem becomes increasingly hard to handle numerically}.  The results concerning the general behavior and drift inversion do 
not significantly depend on the choice of $V_{\rm max}$.  With these parameter choices the relaxation times $\tau$ for $N=3...11$ 
fall between $\omalog{\tau}=0.8,...,4.8$ for RD polymers and $\omalog{\tau}=0.3,...,3.4$ for FM polymers in all potentials studied 
here.  Below we let $N$ and $T$ vary.  By the limit $T \rightarrow 0$ we mean going to the mean-field stochastic generator with 
the rates given by (\ref{eq:meanfield}).  Overviews of the dynamics of polymers of lengths $N=1,3,5,7,9$ are given in Figs.~3 and 
5, while Figs.~4 and 6 provide more detail for $N=1...11$.  We are especially interested in the current inversions and the general 
effects of the polymer size.

\begin{figure}
\includegraphics[width=8.0cm]{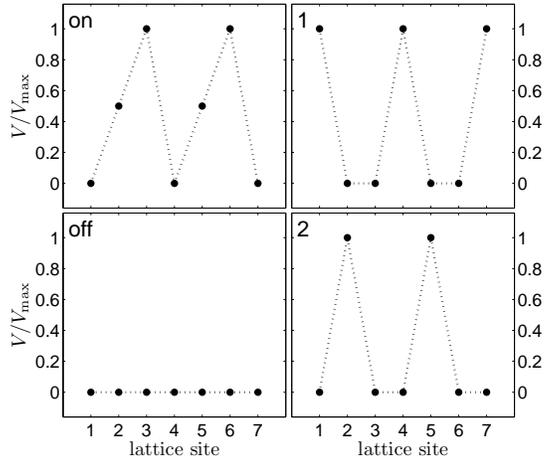}
\caption{Sketch of the flashing non-symmetric ratchet (left columns) and traveling symmetric ratchet (right columns) for $L=3$ 
(two period lengths shown).}
\end{figure}

\subsection{Flashing ratchet potential}

Let us first define the time-period $T=T_{\rm on}+T_{\rm off}$ and the symmetry parameter $x=T_{\rm off}/T$, where $T_{\rm 
on/off}$ are the corresponding mean life-times of the potentials (see Fig.~2).  In previous studies ({\it 
e.g.}~\cite{Schimansky,Zhou}), only symmetric flashing $x=1/2$ was considered.  This results in zero drift for $T \rightarrow 0$, 
which also happens in all real systems (for all $x$).  However, with $x \neq 1/2$, this does not happen for the models considered 
here.  The drift changes its sign as a function of $x$ and the point of this sign change in $x$ depends on $T$.  This is shown in 
Fig.~3, where we have plotted $v$ as a function of $x$ with three different $T$'s (figures (a)-(c)) that represent the general 
behavior in different scales of $T$.  The drift in the positive direction (generated by the short slope) arises when the ratchet 
is switched on for such a short time that the larger rate of the short slope wins the smaller rate of the longer slope (see 
Fig.~2).  Therefore, for increasing $T$, the ratio $x$ must get smaller to retain the dominance of the shorter potential slope, 
and finally $x$ goes to zero at $T \rightarrow \infty$.

When we add more reptons, the overall shape of the $v$ curves remains very similar with small $T$'s.  However, a clear effect of 
the polymer length and internal mechanisms can be seen with the long-time period $T=\omaexp{5}$, where the drift curve of the 
$N=9$ RD polymer turns positive for $x \approx 1/2$.  The velocities of the FM polymers remain on the negative side and no change 
in their general drift behavior can be seen as the parameters $N$ and $T$ are varied.  In Fig.~3 (d) we have plotted an example of 
$D_{\rm eff}$ behavior of the $N=9$ polymers with corresponding $T$'s of the figures (a)-(c).  Here $x=1$ simply gives the 
diffusion coefficient of the free polymers, and the diffusion constant in the static potential (at $x=0$) is always smaller.  As 
can be seen, the effect of $T$ and $x$ on the diffusion is quite small in general.

\begin{figure}
\includegraphics[width=9.0cm]{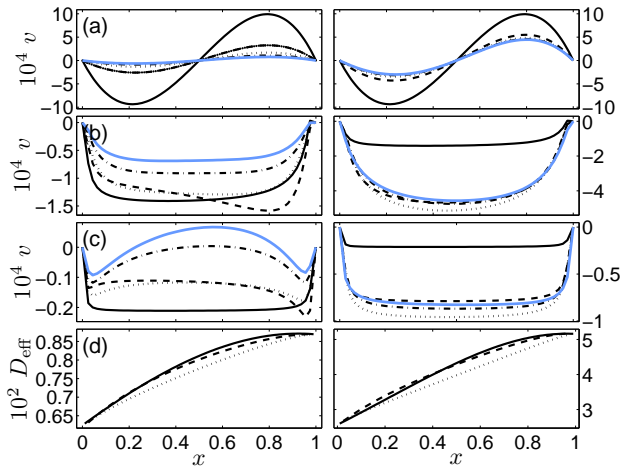}
\caption{(Color online) Drift and diffusion of the RD and FM polymers in the flashing ratchet ($L=3$, $V_{\rm max}=1/2$).  (a-c): 
drift as a function of the symmetry parameter $x=T_{\rm off}/T$ with the total flashing period $T \rightarrow 0$ (a), $T= 
\omaexp{3}$ (b) and $T=\omaexp{5}$ (c) with $N=1$ (solid black), $N=3$ (dash), $N=5$ (dot), $N=7$ (dash-dot), $N=9$ (solid blue).  
(d): effective diffusion coefficient of $N=9$ with $T \rightarrow 0$ (solid), $T=\omaexp{3}$ (dash) and $T=\omaexp{5}$ (dot).}
\end{figure}

Next we fix values $\boldmath{x}=\boldmath{1/4},\boldmath{1/2},\boldmath{3/4} \text{}$ and examine the $T$ dependence of the drift 
and the Peclet number in detail.  The results in Fig.~4 reveal a complex behavior of the drift.  The overall form of the curves is 
as expected: the drift and the Peclet number have some (local) maxima around $\omalog{T} \approx 0$.  For small $T$, the single 
particle remains the fastest in all cases excluding $x=1/2$ for FM polymers, where it is the slowest one.  However, as $T$ gets 
larger, longer polymers eventually become faster, which is caused by their longer relaxation time (short polymers have already 
reached their stationary state).  This can be clearly seen from Figs.~4 (b) and (c), but it also takes place in figure (a) to some 
extent.  Similar behavior of coupled particles being faster than single ones and also having drift inversions were also reported 
in Ref.~\cite{Klumpp}.  Although the relaxation times are quite different (see Sec.~III A), the maxima of the drift fall close to 
$\omalog{T} \approx 1$ for all polymer lengths and the position of the maximum Peclet number is almost constant.  The drift sign 
change, already seen in Fig.~3, is present in Fig.~4 (c).

The behavior of the Peclet number is very clear and similar in every case in Fig.~4: the larger the polymer, the larger the Peclet 
number.  Thus the transport of longer polymers is more coherent than of shorter ones.  Similar behavior was found in a continuum 
model consisting of elastically coupled Brownian particles \cite{Wang}.  By comparing the values of the Peclet number between 
polymer types, we see no significant differences between the curves.  There is a slight difference for large values of $T$, where 
the Peclet number remains larger for FM polymers.  This holds with every choice of parameters, excluding the possible current 
inversion points ({\it e.g.}~the interval $\omalog{T}=-1...0$ in Fig.~4 (c)).

\begin{figure}
\includegraphics[width=9.0cm]{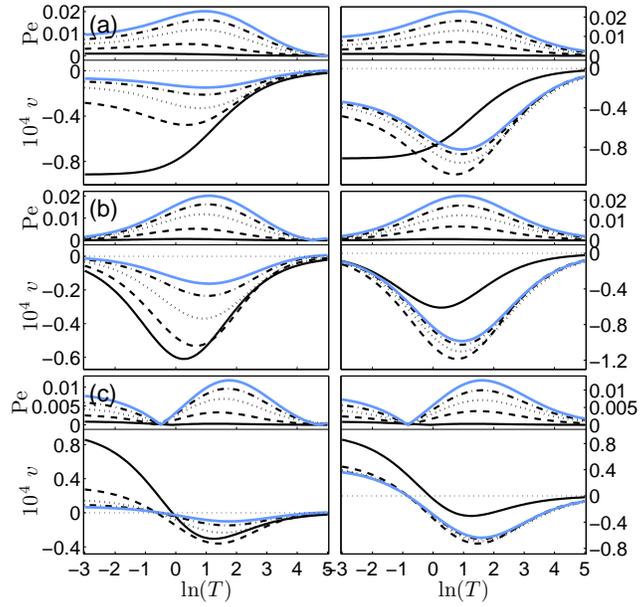}
\caption{(Color online) Drift and Peclet numbers of the RD polymers (left column) and FM polymers (right column) in the flashing 
ratchet ($L=3$, $V_{\rm max}=1/2$) with $N=1$ (solid black), $N=3$ (dash), $N=5$ (dot), $N=7$ (dash-dot), $N=9$ (solid blue).  
Symmetry parameters $x=T_{\rm off}/T$ are $x=1/4$ (a), $x=1/2$ (b) and $x=3/4$ (c).}
\end{figure}
 
Next we take a closer look at the asymptotic behavior at $T \rightarrow \infty$.  In Fig.~5 we have plotted the mean travel 
distance $d$ defined in the Eq.~(\ref{eq:slowlimit}).  For $N=1,2$ there are no bulk-reptons so the mean travel distances of RD 
and FM polymers may differ for $N \geq 3$ only.  The calculation reveals that for long RD polymers ($N>5$, a 'critical length') 
the rebound effect wins ({\it i.e.}~$d>0$) and the polymer starts traveling backwards while the single particle and FM polymers 
are traveling to the expected negative direction.  The rebound effect is also present in FM polymers, but it is not strong enough 
to reverse the drift direction.  For RD polymers with $L>3$ with feasible polymer lengths our Monte Carlo test simulations do not 
display this kind of an anomalous current inversion, suggesting that it may be related to spatial discretization and that 
longer-range interactions (e.g.~stiffness) between reptons need to be introduced to see such inversions for $L>3$.

\begin{figure}
\includegraphics[width=8.0cm]{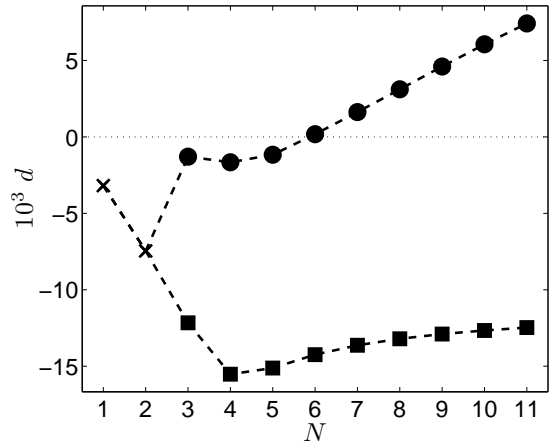}
\caption{Mean travel distances of the RD polymers (circles) and FM polymers (squares) in the flashing ratchet ($L=3$, $V_{\rm 
max}=1/2$) as a function of the polymer length in one time-period at asymptotic limit ({\it i.e.}~the stationary state is reached 
before the switching).}
\end{figure}

We also note that a similar effect of multiple current inversions with tightly connected Brownian particles (rods) was reported in 
Ref.~\cite{Craig2}.  In that work, however, current inversions were not found for objects able to vary their length (rotating 
rods) in the ratchet direction, whereas the polymers in our work are able to vary their length between $1...N$ and still have 
drift inversion.  

The reason for the stronger rebound effect of the RD polymer is caused by the strong tendency to enter (possibly deformed) 
U-shaped configurations because of the strict reptation rule.  After the potential is turned off, this shape unwinds and causes 
the drift.  This also happens with time-dependent fields \cite{Puola2}.  Since FM polymers lack the reptation rule, there is not 
as much variation in their shape as RD polymers have, thus resulting in a weaker rebound effect.

\subsection{Traveling ratchet potential}

Let now $T=T_1+T_2$ for the mean life-times $T_1$ and $T_2$ of the potentials depicted in the right column of Fig.~2 and define 
the symmetry parameter $x=T_{1}/T$.  A Similar drift and diffusion behavior as previously reported in Ref.~\cite{Grier,Pascual} 
for a single particle is expected.  In Fig.~6, we show $v$ as a function of $x$ with three different $T$'s (Figs.~6 (a)-(c)): $T 
\rightarrow 0, \: \omalog{T}=3 \text{ and } \omalog{T}=7$.  The behavior for the single particle is as expected; the drift is 
antisymmetric with respect to $x=1/2$ and goes to zero at $x=0,1/2,1$.  With longer polymers the drift changes sign non-trivially 
for large $T$'s (Fig.~6 (c)) for both polymer types.  This result is unexpected.  An example of the behavior of the diffusion 
coefficient is shown in Fig.~6 (d) for $N=9$ and different $T$'s.  $D_{\rm eff}$ always reaches its maximum at $x=1/2$ and 
decreases as the system goes to a static potential state at $x=0 \text{ and } 1$.  The similarity of Figs.~3 (a) and 6 (a) is 
caused by the fact that, as it can be easily seen from Eq.~(\ref{eq:meanfield}) for $T \rightarrow 0$, the traveling potential 
creates a similar effective rate structure as the non-symmetric flashing ratchet.

\begin{figure}
\includegraphics[width=9.0cm]{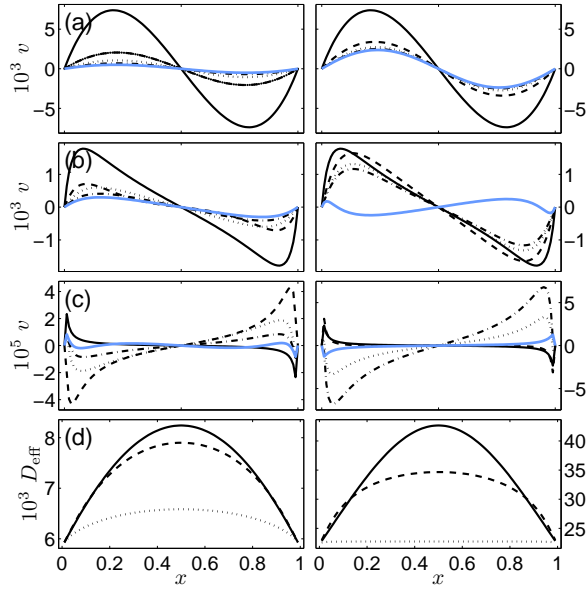}
\caption{(Color online) Drift and diffusion of the RD- and FM polymers in the traveling potentials ($L=3$, $V_{\rm max}=1/2$).  
(a-c): drift as a function of the symmetry parameter $x=T_{1}/T$ with $T \rightarrow 0$ (a), $T=\omaexp{3}$ (b) and $T=\omaexp{7}$ 
(c) with $N=1$ (solid black), $N=3$ (dash), $N=5$ (dot), $N=7$ (dash-dot), $N=9$ (solid blue).  (d): the effective diffusion 
coefficient for $N=9$ with $T \rightarrow 0$ (solid), $T=\omaexp{3}$ (dash) and $T=\omaexp{7}$ (dot).}
\end{figure}

Next we fix $\boldmath{x}=\boldmath{1/4} \text{}$ and examine the $T$ dependence in detail.  In Fig.~7, we have plotted $v$ and 
the Peclet number for $\omalog{T}=-4...7.5$.  As $N>2$, drift inversions can be seen around $\omalog{T} \approx 2$ for both 
polymer types.  As before, the single particle remains the fastest for small $T$, but eventually the drift curves begin to 
intersect as $T$ gets larger and the single particle is not always the fastest (see {\it e.g.}~the $N=3$ FM polymer in Fig.~7 (a), 
right column).  The behavior of the Peclet number is as before: Longer polymers have more coherent transport, excluding the 
possible drift inversion points and their neighborhood.  With small values of $T$, the Peclet number is also the same for both 
polymer types, but because of unequal drifts for moderate and large values of $T$ ($\omalog{T} \geq 0$), also differences exist.

\begin{figure}
\includegraphics[width=9.0cm]{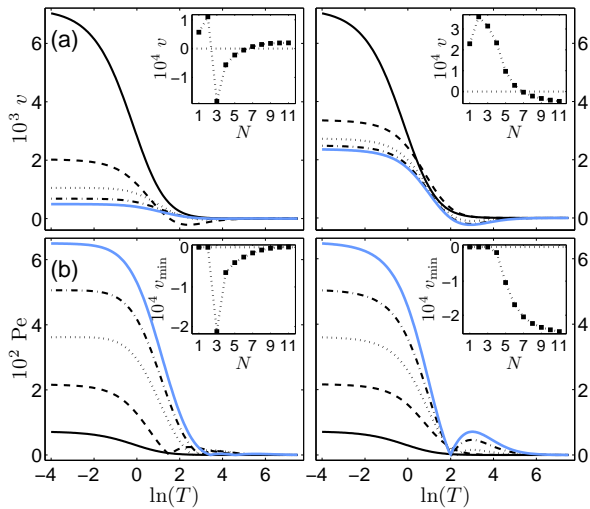}
\caption{(Color online) Drift and Peclet numbers of the RD polymers (left column) and FM polymers (right column) in the traveling 
ratchets ($L=3$, $V_{\rm max}=1/2$) as a function of the mean time-period $T$ with the symmetry parameter $x=T_1/T$ and $N=1$ 
(solid black), $N=3$ (dash), $N=5$ (dot), $N=7$ (dash-dot) and $N=9$ (solid blue).  For the left inset of (a) $\omalog{T}=2.85$ 
and $\omalog{T}=2.05$ for the right inset.}
\end{figure}

The insets of Fig.~7 show the drift as a function of $N=1...11$ in detail.  We have chosen $\omalog{T}=2.85$ for RD polymers and 
$\omalog{T}=2.05$ for FM polymers.  With these choices, the drift inversion occurs between $N=6$ and $7$ for both models.  In the 
insets of Fig.~7 (b) we have plotted the overall drift minimal values in the interval $\omalog{T}=-4...7.5$ as a function of $N$.  
The distinction between the polymer types is very clear.  FM polymers drift increasingly fast backwards whereas RD polymers 
eventually stop moving as $N$ gets larger.  The drift inversion of the RD polymers $N \geq 10$ would require a smaller fixed $x$.  

The magnitude of the drift, typically between $10^{-5}$ and $10^{-3}$, is comparable with the drift caused by a flashing ratchet.  
The Peclet number values of the polymer motion remain small ($ \ll 0.1$) for both potential types, indicating very low coherence 
of transport.

\section{Conclusions}

We studied the ratchet effect with discrete polymer models in time-dependent potentials using the master equation approach.  We 
gave general equations for calculating the effective diffusion coefficient and drift in time-dependent periodic systems.  Using 
these equations, we performed calculations in the flashing and traveling ratchet potentials for short discrete polymers with the 
Rubinstein-Duke model and a relaxed version of this model allowing tube breaking.  We found complex dynamics that results from the 
non-pointlike structure of the polymers by the coupling between the potential and polymer internal states.  By varying the 
potential switching rates, we found non-trivial inversions of the polymer drift direction, which cannot occur with simple 
pointlike non-interacting particles.  We also found that the Peclet number grows as the polymer gets longer and is largely 
independent of the polymer type thus allowing more coherent transport for longer polymers.  The overall polymer dynamics in 
ratchet potentials was found to be very model specific.  The discretization of the problem in this work may be far from many    
real-world applications but, nevertheless, since our model catches the essential characteristics of the Brownian motor system, we 
expect that similar properties could be found in the nano-scale objects that can be described with discrete states instead, such 
as molecular motors with internal structure.  Drift inversions are especially interesting since they facilitate more efficient 
separation methods of molecules.  The next step would be to consider larger $L$ and $N$ and the differences between deterministic 
and stochastic switching \cite{Kauttonen}.

\appendix

\section{Operators in $H$}

The explicit definitions of the operators in Eq.~(1) are
\begin{multline*}
A_{l}(d) = \{ R(l)+L(l) \} \tilde{n}_{0,1,l} - R(l) \tilde{a}_{1,l}^{\dagger}
- L(l) \tilde{b}_{1,l}^{\dagger} \\
+ L(l) \tilde{n}_{A,1,l} - L(l) \tilde{a}_{1,l} + R(l) \tilde{n}_{B,1,l} - R(l) \tilde{b}_{1,l}
\end{multline*}
\begin{multline*}
B_{y,l} = \{ R(l+f(N-1,y))+ L(l+f(N-1,y)) \} n_{0,N-1,y,l} \\ 
- R(l+f(N-1,y)) a_{N-1,y,l}^{\dagger} - L(l+f(N-1,y)) b_{N-1,y,l}^{\dagger} \\
+ L(l+f(N-1,y)) n_{A,N-1,y,l} - L(l+f(N-1,y)) a_{N-1,y,l} \\
+ R(l+f(N-1,y)) n_{B,N-1,y,l} - R(l+f(N-1,y)) b_{N-1,y,l}
\end{multline*} 
\begin{multline*}
M_{i,y,l} = R(l+f(i,y))(n_{A,i,y,l}n_{0,i+1,y,l} + n_{0,i,y,l}n_{B,i+1,y,l} - a_{i,y,l} a_{i+1,y,l}^{\dagger}-b_{i,y,l}^{\dagger} 
b_{i+1,y,l}) \\ 
+ L(l+f(i,y))(n_{0,i,y,l}n_{A,i+1,y,l}+n_{B,i,y,l}n_{0,i+1,y,l} - a_{i,y,l}^{\dagger} a_{i+1,y,l}-b_{i,y,l} b_{i+1,y,l}^{\dagger}) 
\\
+ \Omega R(l+f(i,y))(n_{A,i,y,l}n_{B,i+1,y,l} + n_{0,i,y,l} n_{0,i+1,y,l} - a_{i,y,l} b_{i+1,y,l} - b_{i,y,l}^{\dagger} 
a_{i+1,y,l}^{\dagger}) \\
+ \Omega L(l+f(i,y))(n_{B,i,y,l}n_{A,i+1,y,l} + n_{0,i,y,l} n_{0,i+1,y,l} - b_{i,y,l} a_{i+1,y,l} - a_{i,y,l}^{\dagger} 
b_{i+1,y,l}^{\dagger}) , 
\end{multline*}
where $\Omega = 0$ for RD polymers and $1$ for FM polymers, and
\begin{align*}
\tilde{a}_{1,l} &= c_{l}^{+} a_{1} \: \quad \tilde{a}_{1,l}^{\dagger} = c_{l}^{-} a_{1}^{\dagger} \\
\tilde{b}_{1,l} &= c_{l}^{-} b_{1} \: \quad \tilde{b}_{1,l}^{\dagger} = c_{l}^{+} b_{1}^{\dagger} \\
\tilde{n}_{z,1,l} &= n_{l} n_{z,1} \\
x_{i,y,l} &= n_l \left( \prod_{j=1}^{i-1} n_{g(y,j),j} \right) x_i \\ 
n_{z,i,y,l} &= n_l \left( \prod_{j=1}^{i-1} n_{g(y,j),j} \right) n_{z,i}
\end{align*}
with $x \in  \{ a,b,a^{\dagger},b^{\dagger} \}$, $z \in  \{ A,0,B \}$.  The function $g(y,i) \in \{ A,0,B \}$ gives the state of 
the $i$th bond in the configuration $y$, and the function $f$ 
\begin{gather*}
f(i,y) =  \sum_{j=1}^{i} \bra{\Psi_{y}} n_{A,i}-n_{B,i} \ket{\Psi_{y}}
,\quad 1 \leq i \leq N-1
\end{gather*}
gives the position of the repton $i+1$ in marker-centered coordinates.  The detailed forms of the functions $g$ and $f$ depend on 
the selection of the state basis.


\begin{thebibliography}{9}

 
 \bibitem{Reimann}
 P. Reimann, 
 Phys. Rep. \textbf{57},
 361 (2002).
 
 \bibitem{Linke}
 H. Linke, 
 Appl. Phys. A \textbf{75},
 2 (2002).
 
  \bibitem{Astumian}
 R.D. Astumian,
 Phys. Chem. Chem. Phys. \textbf{9},
 5067 (2007).
  
\bibitem{Wang}
 H. Wang and J.D. Bao,
 Physica A \textbf{374},
 33 (2006).
 
 \bibitem{Klumpp}
 S. Klumpp, A. Mielke, and C. Wald,
 Phys. Rev. E \textbf{63},
 031914 (2001).
 
 \bibitem{Chen}
 H. Chen, Q. Wang, and Z. Zheng,
 Phys. Rev. E \textbf{71},
 031102 (2005).
 
 \bibitem{Gehlen}
 S. von Gehlen, M. Evstigneev, and P. Reimann,
 Phys. Rev. E \textbf{77},
 031136 (2008).
 
  \bibitem{Fendrik}
 A.J. Fendrik, L. Romanelli, and R.P.J. Perazzo,
 Physica A \textbf{368},
 7–15 (2006). 
 
 \bibitem{Retkute}
 R. Retkute and J.P. Gleeson,
 Fluctuation and noise letters \textbf{6},
 3 (2006).    
 
 \bibitem{Streek}
 M. Streek, F. Schmid, T.T. Duong, and A. Ros
 Journal of Biotechnology \textbf{112},
 79–89 (2004).
 
 \bibitem{Craig1}
 M.T. Downton, M.J. Zuckermann, E.M. Craig, M. Plischke, and H. Linke,
 Phys. Rev. E \textbf{73},
 011909 (2006).
  
 \bibitem{Craig2}
 E.M. Craig, M.J. Zuckermann, and H. Linke,
 Phys. Rev. E \textbf{73},
 051106 (2006).
 
  \bibitem{Fisher}
 A. Kolomeisky and M. Fisher,
 Annu. Rev. Phys. Chem. \textbf{58},
 675 (2007).
 
 \bibitem{Jarzynski}
 C. Jarzynski and O. Mazonka,
 Phys. Rev. E \textbf{59},
 6448 (1999).
 
\bibitem{Zhou}
 Y. Zhou and J.D. Bao,
 Physica A \textbf{343},
 515 (2004).
 
 \bibitem{Schimansky}
 J.A. Freund and L. Schimansky-Geier,
 Phys. Rev. E \textbf{60},
 1304 (1999).

\bibitem{Pascual}
 J. Casado-Pascual,
 Phys. Rev. E \textbf{74},
 021112 (2006).

\bibitem{Rubinstein}
 M. Rubinstein,
 Phys. Rev. Lett. \textbf{59},
 1946 (1987).

\bibitem{Duke}
 T.A.J. Duke,
 Phys. Rev. Lett. \textbf{62},
 2877 (1989).
  
\bibitem{Widom}
 B. Widom, J. Viovy, and A. Defontaines,
 J. Phys. I France \textbf{1},
 1759 (1991).
 
  \bibitem{Carlon}
 E. Carlon, A. Drzewinski, and J.M.J. van Leeuwen,
 J. Chem. Phys. \textbf{117},
 2425 (2002).
 
  \bibitem{Puola2}
 P. Pasciak, K. Kulakowski, and E. Gudowska-Nowak,
 Acta Physica Polonica B \textbf{36},
 1737 (2005). 

  \bibitem{Kooiman}
 A. Kooiman and J.M.J. Van Leeuwen,
 J. Chem. Phys. \textbf{99},
 2247 (1993).

  \bibitem{constrait3}
 A. Drzewinski and J.M.J. van Leeuwen,
 Phys. Rev. E \textbf{77},
 031802 (2008).
  
  \bibitem{constrait1}
 M. Pae\ss{}ens and G.M. Sch\"utz,
 Phys. Rev. E \textbf{66},
 021806 (2002).
  
 \bibitem{constrait2}
 A. Drzewinski and J.M.J. van Leeuwen,
 Phys. Rev. E \textbf{73},
 061802 (2006).
 
 \bibitem{Kramers}
 P. H\"anggi, P. Talkner, and M. Borkovec,
 Rev. Mod. Phys. \textbf{64},
 251 (1990).
  
  \bibitem{Sartoni}
 G. Sartoni and J.M.J. van Leeuwen,
 Phys. Rev. E \textbf{57},
 3088 (1998).

  \bibitem{Derrida}
 B. Derrida,
 J. Stat. Phys. \textbf{31},
 433 (1983).
 
 \bibitem{Kolomeisky}
 A. Kolomeisky,
 J. Chem. Phys. \textbf{115},
 7253 (2001).

 \bibitem{Maritan}
 G. Lattanzi and A. Maritan,
 Phys. Rev. E \textbf{64},
 061905 (2001).
 
 \bibitem{Elston}
 H. Wang and T.C. Elston,
 J. Stat. Phys. \textbf{128}, 
 35 (2007).

 \bibitem{Stukalin}
 E. Stukalin and A. Kolomeisky,
 J. Chem. Phys. \textbf{124},
 204901 (2006).
 
\bibitem{Grier}
 S.H. Lee and D.G. Grier,
 Phys. Rev. E \textbf{71},
 060102(R) (2005).

 \bibitem{Kauttonen}
 J. Kauttonen, J. Merikoski, and O. Pulkkinen, 
 unpublished.
 
 \end{thebibliography}
\end{document}